\documentclass[apjl, iop]{emulateapj} 
\usepackage{amsmath}
\usepackage{amssymb} 
\usepackage{amstext} 
\usepackage{apjfonts}
\usepackage{graphicx} 

\usepackage{color}

\begin{document}

\shorttitle{Avoiding GC disruption through BHXBs} \shortauthors{Justham,
Peng \& Schawinski} \submitted{}

\title{Globular cluster formation efficiencies from black-hole X-ray
  binary feedback}

\author{Stephen Justham}
\affil{The Key Laboratory of Optical Astronomy, National Astronomical
Observatories, The Chinese Academy of Sciences, Datun Road, Beijing
100012, China; sjustham@nao.cas.cn}

\author{Eric W. Peng} 
\affil{Department of Astronomy, Peking University,
Beijing 100871, China \\ The Kavli Institute for Astronomy and
Astrophysics, Peking University, Beijing 100871, China}

\author{Kevin Schawinski} 
\affil{Institute for Astronomy, ETH Zurich, Wolfgang-Pauli-Strasse 27,
  8093 Zurich, Switzerland}

\begin{abstract} 
We investigate a scenario in which feedback from black-hole X-ray
binaries (BHXBs) sometimes begins inside young star clusters 
before strong supernova feedback. 
Those BHXBs could reduce the gas fraction inside embedded young
clusters whilst maintaining virial equilibrium, which may help globular
clusters (GCs) to stay bound when supernova-driven gas ejection subsequently occurs.
Adopting a simple toy model with parameters guided by
BHXB population models, we produce GC formation efficiencies
consistent with empirically-inferred values. 
The metallicity dependence of BHXB formation could naturally
explain why GC formation efficiency is higher at lower metallicity.
For reasonable assumptions about that metallicity dependence,
our toy model can produce a GC metallicity bimodality in some galaxies
without a bimodality in the field-star metallicity distribution.  
\end{abstract}


\section{Introduction} 
\label{sec:intro}

The formation of bound Globular Clusters (GCs) takes place in an
exceptionally complicated environment. A 
complete explanation for observed GC populations likely involves
multiple physical processes
\citep[see, e.g.,][]{SPZ+McMillan+Gieles2010,Longmore+2014},
probably including galaxy merger histories 
\citep[see, e.g.,][]{Kruijssen2014review}.
However, a common simple picture is that proto-clusters become unbound
when supernova feedback (SNF) ejects the remaining cluster gas, if the gas
fraction at that time is too high ($\gtrapprox 0.7$). 
If true, that might suggest that that GC formation is linked
to regions of very high star-formation efficiency ($\gtrapprox 0.3$).

It has been argued that young clusters lose
their gas too early to be consistent with that picture \citep[see,
e.g.,][and references therein]{Longmore+2014}. However, the 
inferred ages of known gas-free clusters which are most similar to
young GCs are still broadly consistent with the model we present
\citep[][]{Bastian+2014}, especially given the likely uncertainty in
determining their ages \citep[see, e.g.,][]{Schneider+2014}. Nonetheless,
if proto-GCs are shown to always lose their gas before SNF occurs 
then our model would not work.

In this Letter we examine whether X-ray binary (XB) feedback (XBF)
might affect the survival probabilites of
young star clusters, since black-hole (BH) XBs
should sometimes form before significant SNF occurs \citep[see][]{JS2012}.
BHXBs may decrease the gas fraction inside the
proto-cluster, and so could increase the effective star formation
efficiency before SNe eject the remaining gas. 

\begin{centering} 
\begin{figure*} 
\begin{centering}
\includegraphics[width=17cm]{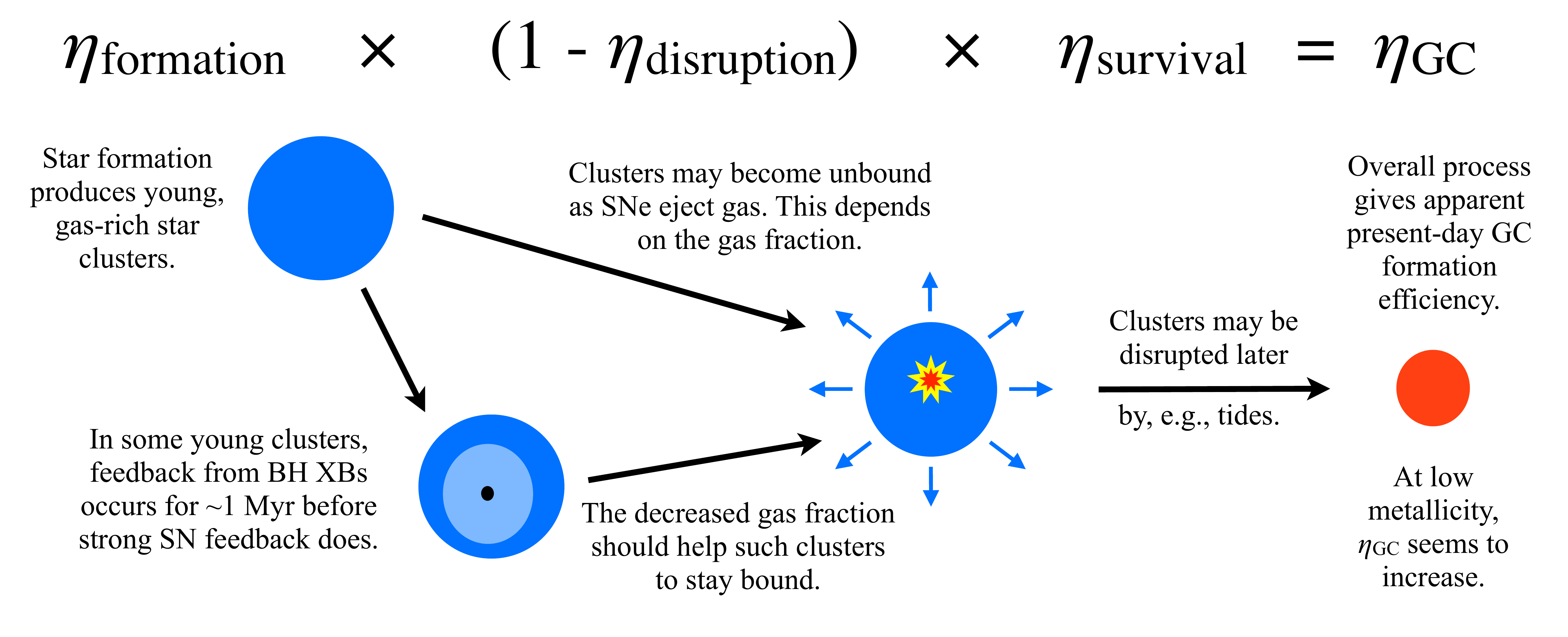}
\caption{\label{fig:schematic} 
The overall formation efficiency of present-day GCs ($\eta_{\rm GC}$) might be
determined by multiple processes. If $\eta_{\rm
GC}$ is Z-dependent, as observations indicate, this may mostly be
contained within $\eta_{\rm disruption}$,
which we suggest XBF may explain. In addition, or alternatively,
$\eta_{\rm formation}$ could be Z-dependent, perhaps indirectly (if,
e.g., low-Z star formation tends to happen at higher intensity or
density). Strong Z-dependence of $\eta_{\rm survival}$ seems harder; it
might rely on the different locations of the Z-rich and Z-poor cluster
populations, since their ages are similar.} 
\end{centering}
\end{figure*} 
\end{centering}

XBF has already been suggested to be especially
important in the epoch of galaxy formation \citep[see,
e.g.,][]{Glover+Brand2003,Power+2009,Cantalupo2010,Mirabel+2011,JS2012,Artale+2015}.
Not only might the radiative feedback from XBs be significant
\citep{Cantalupo2010,Jeon+2012,Power+2013}, so might their kinetic
output \citep{Fender+2005,Heinz+Grimm2005}.  Some XBs
directly drive very energetic kinetic outflows \citep[see, e.g.,][]{Gallo+2005,Pakull+2010,Soria+2014}. 
Stochastic galaxy-to-galaxy variation in XBF may help to explain some
of the diversity in dwarf galaxy populations \citep{JS2012}. 

Here we argue that the influence of BHXBs on the survival of 
proto-GCs might explain why the GC formation
efficency -- $\eta_{\rm GC}$, i.e., the fraction of stellar mass which
stays in bound clusters --  apparently increases at lower metallicity
\citep[see, e.g.,][]{Harris+Harris2002,Forte+2005}. Such a change in
$\eta_{\rm GC}$ is necessary to explain why the GCs in massive,
metal-rich galaxies are predominantly metal-poor \citep[][]{Peng+2008}.
The present-day observed $\eta_{\rm GC}$
presumably represents both formation-epoch processes and subsequent cluster
disruption (see, e.g., \citealt{GnedinOstriker1997} and Figure
\ref{fig:schematic}). Unless the other processes which affect
  $\eta_{\rm GC}$ are strongly metallicity
dependent then the model we present should be able to accommodate them.

We do not examine the formation of multiple stellar
populations in GCs. \citet{Leigh+2013} have done so
when considering single stellar-mass BHs which accrete from the
proto-cluster gas. If BHXBs can affect the formation epoch of GCs, then
they might somehow help to resolve this 
problem.\footnote{By contrast, as we were
finalising this manuscript, \citet{Cabrera-Ziri+2015} speculated that
XBs might have ejected gas from some young clusters, thereby
reducing the opportunity for second-generation star formation.}

Section \ref{sec:input} presents the background to this scenario. Sections
\ref{sec:toy} and \ref {sec:bimodality} investigate the idea using 
  toy models.

\section{Feedback from X-ray binaries in embedded young clusters}
\label{sec:input}

A key property of BHXBs for proto-GC survival is that BHXBs can
be formed before SNe eject the cluster gas. This is possible if
BHs can form relatively quietly, from stars more massive
than those which dominate the energetic SNF 
\citep[see, e.g.,][and refs.\ therein]{Heger+2003,JS2012}.

\citet{RPP2005} find young BHXBs turn on $\approx$4 Myr after a
starburst; their models match the observed luminous XB
distribution in the Cartwheel galaxy. The expected time available for
such BHXBs to act before strong SNF depends on whether  
BH formation ``by fallback'' is sufficiently quiet to avoid sudden gas
ejection, or only formation of BHs by direct collapse. Those options roughly
correspond to initial single-star masses above $\approx 25 M_{\odot}$
and $\approx 40 M_{\odot}$, respectively (a more complete picture would
be metallicity-dependent -- see, e.g., \citealt{Heger+2003} -- and
binarity adds further complexity). 
Nonetheless, by the time SNe explosively eject
gas from the proto-GC, one or more BHXBs might well have had
$\approx$1--5 Myr to reduce the gas fraction inside the cluster, which is a
signficant qualitative difference between those proto-clusters in which
an early BHXB does form and those in which one does not \citep{JS2012}.

There are exceptions to the above broad statements.
For example, \citet{BelczynskiTaam2008} predict that $\sim$0.1
per cent of NSs form within 5 Myr of star formation due to binary
interactions, presumably in energetic SNe. Even rarer are BH-forming
events probably associated with
long-duration gamma-ray bursts \citep[see, e.g., ][and references
therein]{PhP+2004}, and classes of
``super-luminous'' supernovae \citep[see, e.g.,][]{Gal-Yam2012}. 
Our toy models neglect these unusual events, although they may be
sufficiently common to affect detailed cluster demographics.

Another complication is the possibility of age spreads larger
  than $\approx$1 Myr within cluster stellar populations.  Since the
  expectation is that denser clusters should have smaller age spreads
  \citep{Elmegreen2000,TanKrumholtzMcKee2006,Parmentier+2014}, this
  may make gas
  ejection by BHXBs relatively less important for less dense clusters than
  for proto-GCs.

\subsection{Energetics and binding energy} 
\label{sec:energetics}

For a BHXB to significantly alter the distribution of gas within
a protocluster, the energy input
must be at least comparable to the binding energy of the gas.
Adopting a binding energy $E_{\rm bind}$ for that gas of order $G M
M_{\rm gas}/R$=$GM^{2} f_{\rm gas}/R$, 
(where $f_{\rm gas}=M_{\rm gas}/M$ is the fraction of the total 
cluster mass -- $M$ -- in gas, and $R$ is a characteristic radius for
the cluster) we estimate the binding energy of the gas as: 

\begin{equation} 
E_{\rm bind} \sim 10^{50} ~{\rm erg}  \left( \frac{M}{\rm 10^{5}
    ~M_{\odot}} \right)^2 \left( \frac{f_{\rm gas}}{1}
\right) \left( \frac{R}{\rm 10~pc} \right)^{-1} ~~. 
\end{equation}

\noindent Using those characteristic values for $M$, $R$ and $f_{\rm
gas}$ leads to a ratio between $E_{\rm bind}$ and $E_{\rm XBF}$ -- the
energy input from XBF of luminosity $L_{\rm XBF}$ over time $t_{\rm
XBF}$ -- of order: 

\begin{equation} 
\frac{E_{\rm bind}}{E_{\rm XBF}}
\sim \frac{1}{30} \left( \frac{L_{\rm XBF}}{\rm 10^{38} erg~s^{-1}}
\right) \left( \frac{t_{\rm XBF}}{\rm 1~ Myr} \right) 
\end{equation}

\noindent which suggests that a single luminous XB could
significantly affect the gas within a young cluster, even 
for feedback efficiencies as low as a few percent.

\subsection{Column Density}

Another argument in favour of the importance of XBF in young GCs
is that the gas-rich environments of embedded clusters might be 
Compton-thick. This would help to trap the accretion luminosity. 
We take a column density of  $\rm
10^{24}~cm^{-2}$ as sufficient for Compton-thickness, 
or a surface density $\Sigma_{\rm H}$ of slightly above $\rm
1~g~cm^{-2}$. Scaling to this value, we find: 

\begin{equation} 
\left( \frac{\Sigma_{\rm H}}{\rm 1~g~cm^{-2}} \right)
\sim \left( \frac{M_{\rm H}}{10^{6} M_{\odot}} \right) \left(
\frac{R_{\rm GC}}{\rm 10~pc} \right)^{-2} 
\end{equation} 

\noindent where $M_{\rm H}$ is the hydrogen gas mass.  
So for gas fractions of $\approx$ 90\%
(99\%), proto-clusters with initial \emph{stellar}
masses above $\approx 10^{5} M_{\odot}$ ($10^{4} M_{\odot}$) may well be
Compton-thick.    

\citet{Zezas+2002} reported six heavily obscured luminous X-ray sources
 consistent with being inside dense, gas-rich, star-forming clouds in the
  Antennae. Those may well represent less extreme examples of
  such embedded systems.

\subsection{BHXB formation probabilities and metallicity dependence}
\label{sec:poisson}

For studying GC survival, we only consider the most rapidly-formed BHXBs --
those which switch on before strong SNF. Here we assume that the formation of
those pre-SNF BHXBs can be treated as a Poisson process. 

At solar metallicity we take the mean population from the calculations
of \citet{RPP2005}, using the predicted 
number of BHXBs more luminous than $\rm 10^{39}~erg~s^{-1}$ at an
age of $\approx$5 Myr.  At
that time, for a population which will eventually produce
$10^{6}$ core-collapse SNe, their least optimistic model predicts
$\approx$1 such BHXB, and their more optimistic models $\approx$10 of
them. We adopt an intermediate value of 3 (simply normalised to $100 M_{\odot}$
of stars per core-collapse SN), so models allow at least half an
order-of-magnitude of freedom in either direction in this BHXB
frequency.  
 
At ``low metallicity'' -- roughly below 0.1 $\rm Z_{\odot}$ --  we 
assume that the incidence of suitable luminous young BHXBs is 10 times higher than at $\rm
Z_{\odot}$.  Observational evidence of a similar increase 
has become increasingly convincing \citep[see,
e.g.,][]{Mapelli+2010,Prestwich+2013,Brorby+2014,Douna+2015}, and 
an increase was strongly expected by models (e.g., \citealt{Belczynski+2004,Dray2006,
  Linden+2010, JS2012, Fragos+2013Letter}). 
This metallicity dependence likely does not only arise from a change
in which single stars would form BHs, since
other effects are important in forming a system containing a BH in a
close binary.  In particular, early envelope loss -- with respect to the nuclear
evolution of the stellar core -- can lead a massive star
to leave a NS remnant rather than a BH \citep[see,
e.g.,][]{Wellstein+Langer1999,Brown+Lee2004,BelczynskiTaam2008}. 
However, at sufficiently low metallicity, massive stars burn helium before expanding to become
giants, hence their common-envelope phases tend to
happen later in their nuclear evolution, hence they are more likely
to form BHs in close binaries.  This effect may produce a
sudden increase in suitable BHXB formation below the appropriate
threshold metallicity (which might be significant for \S \ref{sec:bimodality}). 
Observational results from \citet{Prestwich+2013} and especially \citet{Douna+2015}
also suggest a sharp increase in the frequency below a
transition metallicity (below $\rm 12+log(O/H) \approx 8$
for the calibrations used, though conversion to an absolute
metallicity scale is non-trivial).

\begin{centering} 
\begin{figure} 
\includegraphics[width=8cm]{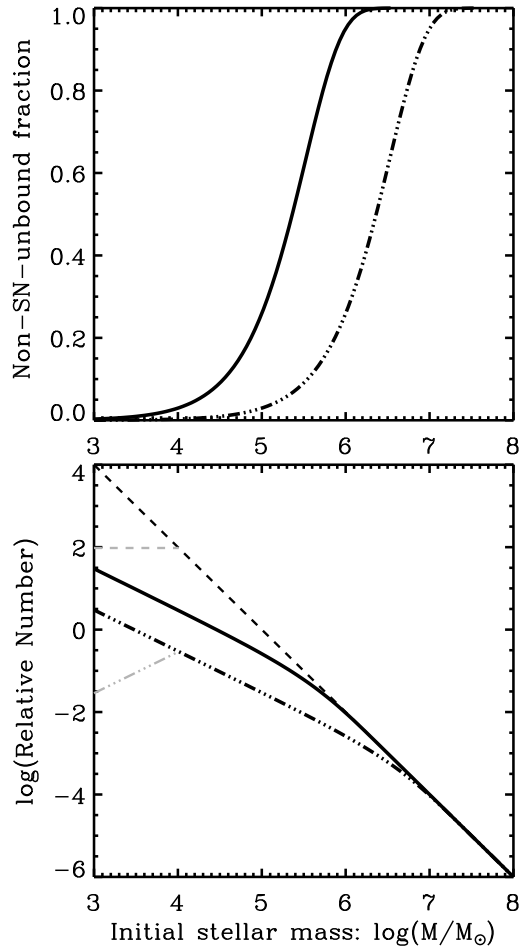}
\begin{centering} 
\caption{\label{fig:survival} The upper panel shows the fraction of
clusters which remain bound in our toy model (dash-dotted
curves represent a $\sim \rm Z_{\odot}$ population, solid curves
represent the low-metallicity cluster population). The lower panel shows
how those survival efficiencies modify a simple power-law ICMF, which is
shown using a dashed line. Using light grey lines, the lower panel
also illustrates the outcome if the ICMF becomes flat below $10^{4}~M_{\odot}$ of
stars.} 
\end{centering}
\end{figure} 
\end{centering}

\begin{centering} 
\begin{figure} 
\includegraphics[width=8cm]{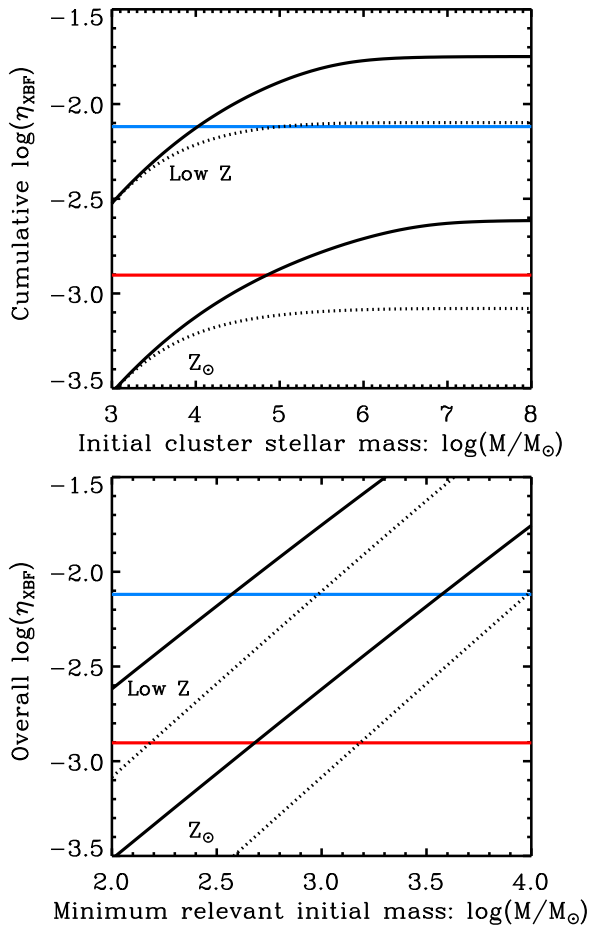}
\begin{centering} 
\caption{\label{fig:efficiency} 
The upper panel shows the cumulative
value of $\eta_{\rm XBF}$ for a toy model when integrating over an
ICMF with a lower-limit of $10^{3} M_{\odot}$.  The lower panel
demonstrates how the overall, integrated $\eta_{\rm XBF}$ is affected by
the lower mass limit of a power-law ICMF. Both panels show
curves for $\sim Z_{\odot}$ and ``low-$Z$'' populations, as indicated.
Solid curves assume the ICMF has a power-law slope of $-2$, as in
Fig.\ \ref{fig:survival}; broken curves adopt an ICMF slope of
$-2.5$. 
Horizontal coloured lines represent 
inferred $\eta_{\rm GC}$ values for red and blue GCs
in NGC 1399 \citep{Forte+2005}.  The 
abscissa in the lower panel is not necessarily the minimum mass of
the ICMF; it may approximate where the ICMF flattens from a power-law.} 
\end{centering}
\end{figure} 
\end{centering}

\begin{centering} 
\begin{figure} 
\includegraphics[width=8cm]{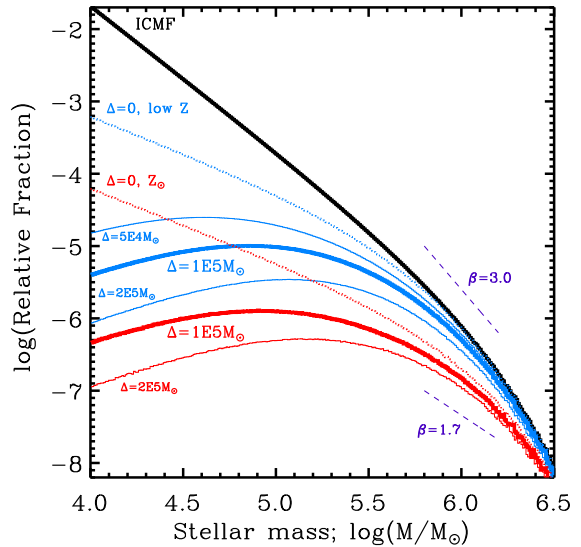}
\begin{centering} 
\caption{\label{fig:withDelta} 
The black solid histogram represents a Schechter-function ICMF,
with an exponential cutoff scale of $10^{6} M_{\odot}$.
The dotted curves (labelled $\Delta=0$) represent model
predictions immediately after gas ejection. The solid coloured histograms
present the mass functions after later mass loss
($\Delta$, as labelled). The dashed purple lines show inferred slopes
for metal-rich and metal-poor cluster populations from \citet{Jordan+2007}.} 
\end{centering}
\end{figure} 
\end{centering}

\begin{centering} 
\begin{figure*} 
\begin{centering}
\includegraphics[width=17cm]{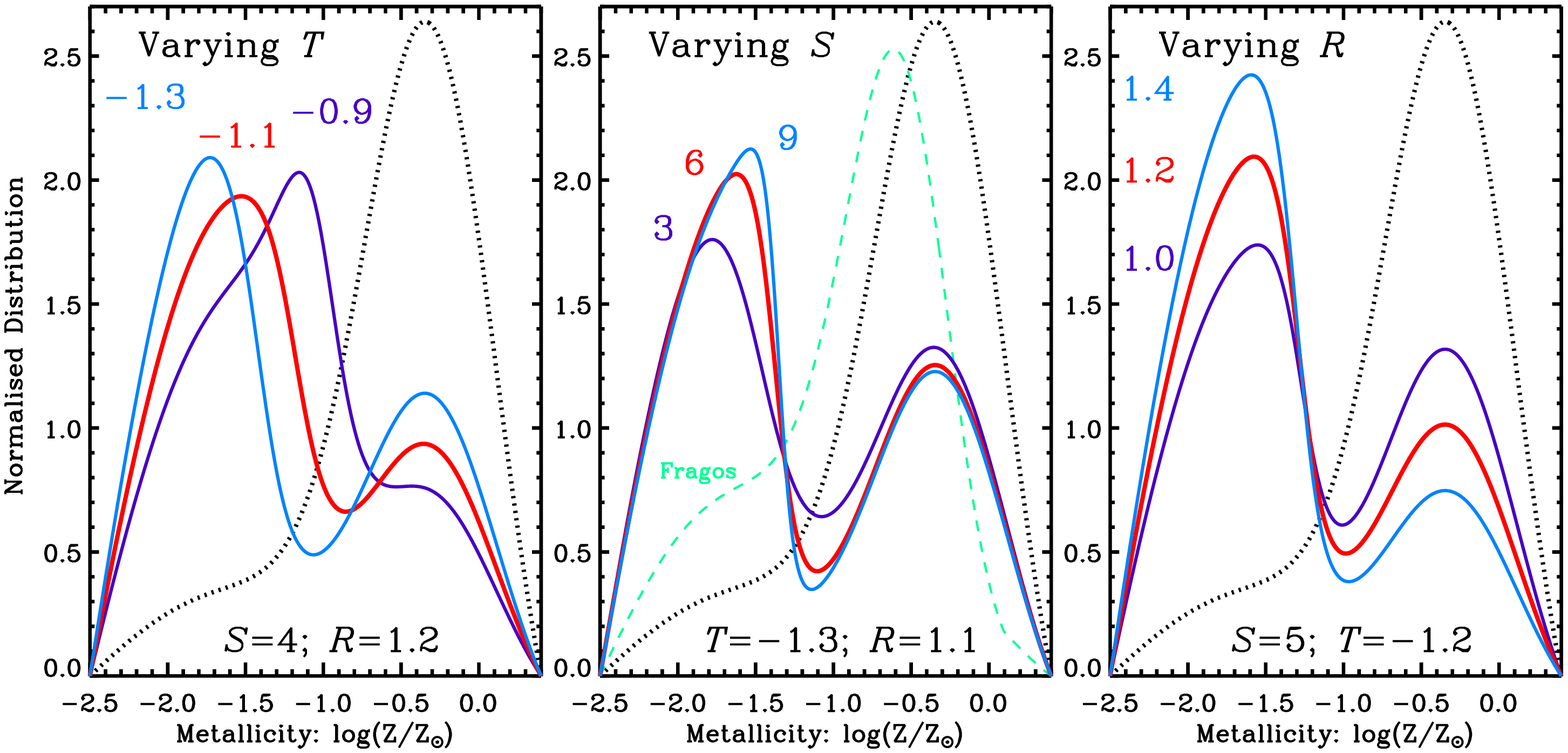}
\includegraphics[width=17cm]{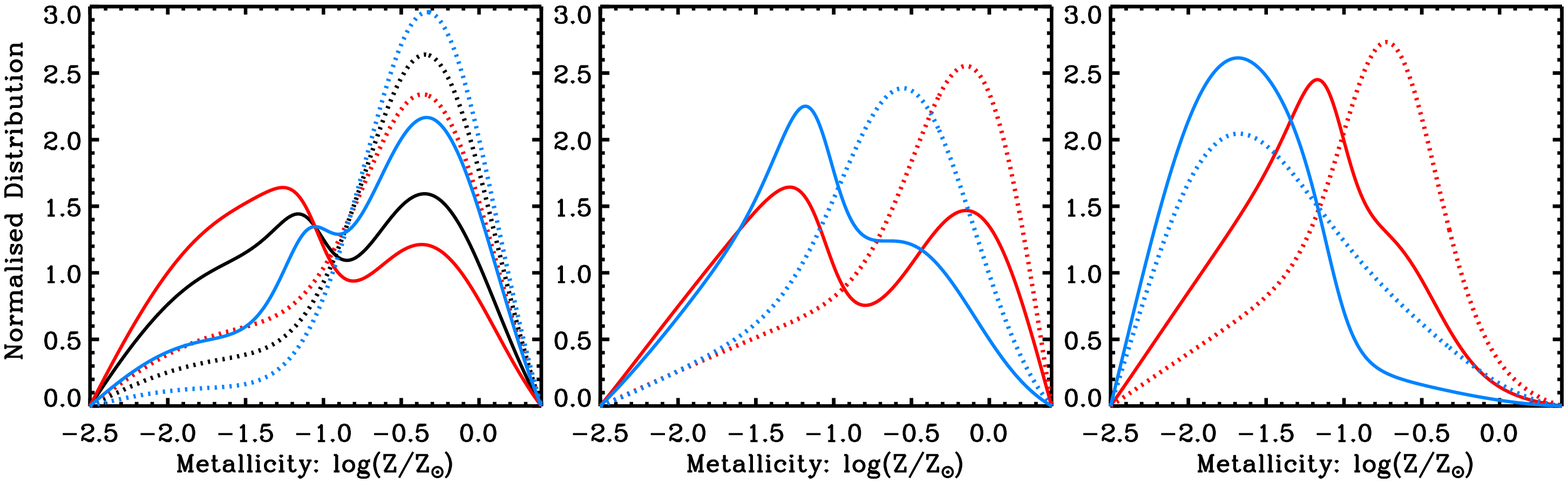}
\caption{\label{fig:bimod} 
For given star formation $Z$-distributions
($\mu (Z)$; the dotted curves), we demonstrate how a toy
model may recover a bimodal distribution in cluster
metallicity (solid curves). \emph{Upper row:} The fixed $\mu (Z)$ is intended to be
broadly representative of common $\mu (Z)$ distributions.  Each
panel varies one of the model parameters, with different 
curves labelled with the relevant parameter value.  The
dashed curve in the center
panel shows the outcome when applying the metallicity dependence
for generic HMXBs from \citet{Fragos+2013Letter}.  \emph{Lower row:}
For fixed parameters ($R=0.7$, $S=5$, $T=1$), we show results
from varying $\mu (Z)$; in the left-hand panel, the black curves are
for the same $\mu (Z)$ as in the top row.} 
\end{centering} 
\end{figure*} 
\end{centering}

\section{Survival probabilities and formation efficiencies} 
\label{sec:toy}

We now use toy models to demonstrate potential consequences of this
scenario.  These adopt a very simple set of assumptions, including
that star formation efficiency is so low that SN-driven ejection of
the remaining gas would normally unbind a young cluster. So a
cluster survives to become a GC \emph{if and only if} a BHXB acts first. 
In the terminology of Fig.~\ref{fig:schematic}, we assume that all
stars form in clusters ($\eta_{\rm formation}=1$; though some clusters may
have very low mass), and that if a BHXB is formed before SNF the cluster
remains bound ($\eta_{\rm XBF}=(1-\eta_{\rm disruption})$).

\subsection{Toy model without later evolution}

First we neglect later disruption processes ($\eta_{\rm XBF}=\eta_{\rm
  GC}$). Then then probability of a cluster remaining
bound is the Poisson probability
of that mass of stars forming one or more BHXBs before SNF (taken from
\S \ref{sec:poisson}). In the statistical limit, this probability is also the fraction of
clusters which remain bound with that particular initial mass (and metallicity).

Figure \ref{fig:survival} presents outcomes from such
  probability calculations.
The upper panel shows survival
fractions for given initial cluster masses, and the lower panel
the effect on a population of clusters with a given initial
cluster mass function (ICMF).
The ``initial cluster masses'' are stellar masses, not including
gas. Since the survival fraction is very low below the mass
range which typically produces GCs we conclude that, if this toy model is
inappropriate below the mass range which produces GCs, it
would only moderately affect broad predictions.

Figure \ref{fig:efficiency} compares integrated survival efficiencies
to present-day inferred $\eta_{\rm GC}$ values
\citep[from][]{Forte+2005}.  The upper and lower panels demonstrate the effect of
varying the uncertain upper and lower bounds of the integral over the ICMF.  
The values predicted for $\eta_{\rm GC}$, along with the difference
between $\eta_{\rm GC}$ for metal-rich and metal-poor clusters, appear
at least order-of magnitude consistent with observationally-derived
values in most of the parameter space.

Where the predicted $\eta_{\rm GC}$ appears too high, this allows
freedom for unclustered star formation and additional GC disruption
mechanisms.  So it seems preferable if the
ICMF flattens or truncates at a few hundred solar masses or more. 
This constraint would relax slightly if suitable BHXBs are more common than we have
assumed; we repeat that in BHXB models there is at least half an
order-of-magnitude of freedom to allow this.

\subsection{Toy model including later evolution}

To our earlier assumptions we now add a
parameterized version of later cluster evolution (due to
\citealt{Fall+Zhang2001}, via \citealt{Jordan+2007}).
For this we use a Monte Carlo approach, randomly drawing
clusters from a Schechter-function ICMF and deciding whether each
cluster remains bound based on the
same BHXB formation probabilities as above. 
Each of the surviving clusters is then assumed to lose a
mass -- denoted by $\Delta$ -- after the
formation epoch \citep[][]{Fall+Zhang2001,Jordan+2007}.

Due to the large
dynamic ranges, for the distributions in Figure
\ref{fig:withDelta} we drew $10^{11}$ primordial clusters from the
ICMF. Over the mass range in Fig. \ref{fig:withDelta}, assuming
$\Delta=10^{5} M_{\odot}$ reduces $\eta_{\rm GC}$ by a factor of a
few. This reproduces the GCMF turnover within our earlier
  model uncertainties.

\subsection{The upper slope of the cluster luminosity function}

\citet{Jordan+2007} found that the bright end of the GC luminosity
function is steeper for GC populations around less massive host
galaxies (which correlates with more metal-poor GC populations).
Figure \ref{fig:withDelta} displays the power law index ($\beta$) inferred by
\citet{Jordan+2007} for the upper mass function of those presumed metal-rich and
metal-poor populations  ($\beta=1.7$ and $3.0$, respectively). 
The correspondence with this model seems at least intriguing.

This qualitative effect also arises in our
  simplest toy model, as Fig.\ \ref{fig:survival} predicts
that the upper mass function of bound GCs should be steeper for
low-metallicity clusters.

\section{Cluster metallicity distributions}
\label{sec:bimodality}

If the formation of appropriate BHXBs has a suitable
  metallicity dependence, then this scenario might
explain the metallicity bimodality of Galactic GCs. 
It is not certain whether the
commonly-observed colour bimodality of extragalactic GCs indicates a
similar metallicity bimodality \citep{YoonYiLee2006,Blakeslee+2012}.
A metallicity-dependent $\eta_{\rm  GC}(Z)$  would produce a GC
metallicity bimodality when the host galaxy's stellar metallicity distribution
$\mu (Z)$ was appropriate -- i.e.\ only if $\mu(Z) \times \eta_{\rm GC}
(Z)$ leads to a bimodal distribution.  Unfortunately, robust $\mu
  (Z)$ distributions for entire galaxies, including their
  halo populations, are difficult to obtain.

Here we show that a simple $\eta_{\rm GC} (Z)$ function, guided by our
model, could reproduce 
a GC metallicity bimodality from a non-bimodal $\mu (Z)$.
We assume two regimes of constant $\eta_{\rm
GC}$ -- one at low $Z$ and one around $Z_{\odot}$ -- matched by a
transition region, expressed as:

\begin{equation} 
\label{eq:tanh_function} 
log \left( \frac{ \eta_{\rm GC} (Z)}{\eta_{\rm GC} (Z_{\odot})} \right) = \frac{R}{2} \left( 1 +
tanh\left( S \left[ T - log \left( \frac{Z}{Z_{\odot}} \right) \right]
\right) \right) 
\end{equation} 

\noindent where $R$ represents the amount
by which $log(\eta_{\rm GC})$ increases at low-$Z$, $S$
gives the ``sharpness'' of the transition between the  $Z_{\odot}$ and
low-$Z$ regimes, and $T$ indicates the metallicity around which that
transition occurs.

For the upper row of Figure \ref{fig:bimod}, we assume a
$\mu (Z)$ constructed to approximate the $\mu (Z)$
distribution of NGC 1399 in figure 5 of \citet{Forte+2005}.  
There we keep $\mu (Z)$ fixed, and show cluster
metallicity distributions for different values of $R$,
$S$ and $T$. 
Our adopted values of $R$ and $T$ are similar to those
which we have previously suggested are expected (i.e., $\approx1$ and $\approx-1$,
respectively; see \S \ref{sec:poisson}). From this $\mu (Z)$ we can
recover a bimodality in cluster
metallicity, and the qualitative result is not affected by
small variations in $S$. The gradual metallicity
dependence from \citet{Fragos+2013Letter} for
\emph{generic} high-mass XBs would not produce a bimodality, as shown in the
central panel. However, the systems involved in our scenario are a very
specific subset of young BHXBs. Moreover, there are 
  indications that a sharp transition is at least plausible
  (see \S~\ref{sec:poisson} and \citealt{Douna+2015}).

The bottom row of Figure \ref{fig:bimod} demonstrates
  changing $\mu (Z)$ with fixed model
  parameters ($R=0.7$, $S=5$, $T=1$).  This
illustrates that the model does not predict that every galaxy
must have a bimodal cluster metallicity distribution \citep[for which
see, e.g.,][]{Usher+2012}, and that the
peaks of the GC metallicity distribution are affected by the underlying
$\mu (Z)$.

When future observations provide reliable $\mu (Z)$
  distributions for a significant sample of galaxies, this should
  enable strong constraints on our
  model.  If Eq.\,\ref{eq:tanh_function} provides a good functional form for
$\eta_{\rm GC} (Z)$ -- and if $\eta_{\rm GC}$ is \emph{only} a function
of metallicity -- then the parameters $R$, $S$ and $T$ should be
universal constants. However, we expect that reality is more
complicated. For example, this simplified model assumes that $\eta_{\rm XBF}$ -- and
hence $\eta_{\rm GC}$ -- has only two regimes ($Z_{\odot}$ and low-$Z$). At
higher metallicities $\eta_{\rm XBF}$ might well decrease
further.  Indeed, for metallicities significantly above solar then
single stars might no longer leave BH remnants at all \citep[see,
e.g.,][]{Heger+2003}.

\section{Proton-capture reactions and abundance anticorrelations} 
\label{sec:protoncapture}

Since the Na-O and Mg-Al anticorrelations are exclusively
associated with star clusters which stay bound, it would be 
elegant if these anticorrelations could be explained by the BHXBs
which -- in this model -- allow clusters to stay bound.  
This speculation arose since the relevant nuclear
reactions are high-energy proton capture reactions (see, e.g.,
\citealt{Izzard+2007}), and many BHXBs may be prolific
sources of high-energy protons
\citep[see, e.g.,][]{Heinz+Sunyaev2002,Fender+2005}.

Unfortunately, even for assumptions which we expect are extremely
  optimistic, we have been unable to convince ourselves that a single
  BHXB would explain the observed abundance anomalies in an entire GC.

Nonetheless, variations on this mechanism may deserve
further examination, perhaps involving other proton accelerators 
in these dense stellar systems.

\section[]{Conclusions}

We have investigated a scenario in which young stellar clusters are
able to survive as bound GCs \emph{only} when feedback from BHXBs 
gradually decreases the gas fraction inside the cluster before energetic
SNe suddenly eject the remainder of the gas. Several
potential complications are ignored by our toy model. However, as
long as the other processes which affect $\eta_{\rm GC}$ are
largely metallicity-independent, this scenario may well have sufficient
freedom to include them without losing its positive
features. A minimal set of assumptions, combined with input
numbers estimated from population models for suitable BHXBs,
produces predictions consistent with observationally-inferred $\eta_{\rm GC}$,
including the increase in $\eta_{\rm GC}$ at low metallicity. We
therefore suggest that more detailed study of early feedback from
BHXBs in proto-GCs may be important for trying to understand GC
populations.

\section*{Acknowledgements}

The authors are grateful to two anonymous referees for enthusiasm and
  comments which led to a much-improved manuscript.
SJ thanks Philipp Podsiadlowski, Youjun Lu and Tassos Fragos for
helpful discussions,
Nathan Leigh and Rainer Spurzem for encouragement, 
Tom Maccarone for wisdom and having a better memory
than him, and the Chinese Academy of
Sciences (President's International Fellowship Initiative Grant No.~2011Y2JB07)
and the National Natural Science Foundation of China (NSFC; Grant
Nos.~11250110055 and 11350110324) for
support. EWP acknowledges support from the NSFC (Grant No.~11173003), and from the
Strategic Priority Research Program, "The Emergence of Cosmological
Structures", of the Chinese Academy of Sciences (Grant No.~XDB09000105).
KS gratefully acknowledges support from Swiss
National Science Foundation Grant PP00P2\_138979/1.   
This research was partially supported by the National Science
Foundation (Grant No.~PHY05-51164) during the KITP program 
``The Formation and Evolution of Globular Clusters''.

\end{document}